\documentstyle[aps,prd,eqsecnum]{revtex}

\begin{document}

\title{
Non-precessional spin-orbit effects
on gravitational waves from inspiraling compact binaries
to second post-Newtonian order
}
\author{
Benjamin J. Owen
\footnote{Electronic address: owen@tapir.caltech.edu}
}
\address{
Theoretical Astrophysics, California Institute of Technology, Pasadena CA 91125
}
\author{
Hideyuki Tagoshi
\footnote{Electronic address: tagoshi@yso.mtk.nao.ac.jp}
}
\address{
National Astronomical Observatory, Mitaka, Tokyo 181, Japan
}
\author{
Akira Ohashi
\footnote{Electronic address: ohashi@tap.scphys.kyoto-u.ac.jp}
}
\address{
Department of Physics, Kyoto University, Kyoto 606-01, Japan
}
\date{\today}
\maketitle

\begin{abstract}
We derive all second post-Newtonian (2PN), non-precessional effects
of spin-orbit coupling on the gravitational wave forms emitted by
an inspiraling binary composed of spinning, compact bodies
in a quasicircular orbit.
Previous post-Newtonian calculations of spin-orbit effects (at 1.5PN order)
relied on a fluid description of the spinning bodies.
We simplify the calculations by introducing into post-Newtonian theory
a $\delta$-function description of the influence of the spins
on the bodies' energy-momentum tensor.
This description was recently used by Mino, Shibata, and Tanaka (MST)
in Teukolsky-formalism analyses of particles orbiting massive black holes,
and is based on prior work by Dixon. 
We compute the 2PN contributions to the wave forms by
combining the MST energy-momentum tensor with 
the formalism of Blanchet, Damour, and Iyer for evaluating
the binary's radiative multipoles, 
and with the well-known 1.5PN order equations of motion for the binary.
Our results contribute at 2PN order only to the amplitudes of the wave forms.
The secular evolution of the wave forms' phase---the quantity most accurately
measurable by LIGO---is not affected by our results until 2.5PN order,
at which point other spin-orbit effects also come into play.
We plan to evaluate the entire 2.5PN spin-orbit contribution to the
secular phase evolution in a future paper,
using the techniques of this paper.
\end{abstract}

\section{Introduction}
\label{s:intro}

Inspiraling compact binaries are one of the main classes
of gravitational wave source to be targeted
by the coming generation of ground-based laser interferometers
such as LIGO, VIRGO, GEO600, and TAMA~\cite{Thorne97}.
There are two reasons for this.
First, binary coalescences are expected to occur fairly often
within detection range of ``enhanced'' interferometers~\cite{LIGO-RD}.
Astronomical lore estimates several neutron-star--neutron-star
coalescences per year within 200~Mpc~\cite{NPS91,Phinney91}
and a similar rate of black-hole--black-hole coalescences within
200~Mpc to 1~Gpc~\cite{Phinney91,TY93,LPP97}.
Second, the signal from the final moments of inspiral
is characterized by a complicated phase evolution
containing detailed information about the physical parameters of the binary,
such as the masses of the bodies and
their spins about their own axes~\cite{Thorne97}.

Because inspiral signals have such a complicated structure,
and because they last many cycles within the frequency bands
of ground-based interferometers,
they are ideal candidates
for the use of matched filtering~\cite{Sathyaprakash}.
Matched filtering,
a signal-processing technique well-studied in the context of radar,
can be used both to search for signals in noisy data and
to estimate parameters once a signal is found.
Matched filtering essentially entails
cross-correlating noisy interferometer data
with a set of theoretical template wave forms.
If a template wave form is a good approximation to the signal wave form,
the cross-correlation enhances the signal-to-noise ratio.
In the context of matched filtering,
a good approximation means (roughly speaking) one in which
the phase evolution of the template matches that of the signal
to within a half cycle out of the total spent
in an interferometer's band.
Because signals are expected to last up to tens of thousands of cycles
in the bands of some interferometers,
the templates must match any possible signal
to a correspondingly high degree of precision.

Currently there is no exact solution to the generic two-body problem
in general relativity.
Thus, inspiral wave-form templates are constructed
using approximation schemes which must be carried out to high precision
to be useful for matched filtering.
These approximation schemes can be broadly grouped into two categories,
the post-Newtonian approach and the black-hole perturbation approach.

The post-Newtonian approach is the longtime standard for
gravitational wave generation. 
It involves expanding the Einstein equations and equations of motion
in powers of the binary's orbital velocity $v/c$
and gravitational potential $GM/rc^2\sim(v/c)^2$,
where the order in $GM/rc^2$ is referred to
as the post-Newtonian (PN) order.
Concurrently, the gravitational wave forms and luminosity
are expanded in terms of time derivatives of
symmetric, trace-free (STF) radiative multipoles,
which are expressed as integrals of the matter source and gravitational fields.
The radiative multipoles are combined with post-Newtonian equations of motion
to yield explicit expressions for the wave forms,
including the secular evolution of orbital phase and frequency
due to radiation reaction.
Recent summaries of the two main versions of the post-Newtonian approach
are given by Blanchet, Damour, and Iyer~\cite{BDI95}
and by Will and Wiseman~\cite{WW96}.
The post-Newtonian expansion of the wave forms of a binary
currently has been carried out to 2.5PN order~\cite{Blanchet96}
[i.e., to $O(v/c)^5$ beyond leading order radiation reaction
and $O(v/c)^5$ beyond Newtonian gravity in all non-radiation reaction effects]
in the case where the two bodies orbit in a quasicircular fashion
and do not spin about their own axes.
(By ``quasicircular'' we mean orbits that are circular
aside from gradual inspiral due to gravitational radiation reaction.)

In the case where the bodies do spin,
there are three types of spin effects to be considered.
Effects of the first type, due to precession of the plane of the orbit,
modulate the amplitude and frequency of the gravitational radiation
in a complicated, non-monotonic fashion.
Secular or dissipative effects, due to radiation reaction, contribute to the
(monotonic) phase and frequency evolution of the orbit;
and non-dissipative effects contribute directly to
the amplitudes of the various harmonics of orbital frequency
in the wave forms, without affecting their phase evolution.
All three types of effects can be further divided into
spin-orbit contributions (i.e., terms involving one spin only)
and spin-spin contributions (interactions between spins).
Precessional effects were first extensively investigated by
Apostolatos {\it et al.}~\cite{ACST94} and by Kidder~\cite{Kidder95}
and were found to complicate matters considerably.
Therefore, like most other treatments of spin,
ours will investigate the case where there is no precession---i.e.,
the spins are parallel or antiparallel to the orbital angular
momentum---leaving precession for future studies.

Non-precessional spin effects have been evaluated
by Kidder, Will, and Wiseman~\cite{KWW93,Kidder95} only to lowest order:
1.5PN for dissipative (and 1PN for non-dissipative) spin-orbit effects,
and 2PN for spin-spin effects.\footnote{
Like non-spin effects, spin effects appear in the
secular phase evolution of the wave forms at a certain order,
and at every order in $v/c$ (0.5PN order) beyond it except for the first.
Thus, spin-orbit effects appear at 1.5PN, 2.5PN, 3PN\ldots\ orders
and spin-spin effects appear at 2PN, 3PN, 3.5PN\ldots\ orders.
}
The main reason for the discrepancy in progress
between the spinning and nonspinning cases
is the form of the matter source used in the Einstein equations.
In the nonspinning case it is simple to write the energy-momentum tensor
as a Dirac $\delta$-function, which greatly simplifies the calculations.
In order to derive spin effects, Refs.~\cite{KWW93,Kidder95}
treated the bodies as uniformly rotating balls of perfect fluid.
The perfect fluid energy-momentum tensor 
was integrated over a finite spatial volume,
which made the multipole integrals much more cumbersome
than in the $\delta$-function case
and introduced additional complications
in the definition of the binary's center of mass.
The net result was that spin calculations at a given post-Newtonian order
seemed to require as much effort as spinless calculations
at higher post-Newtonian order,
and spin calculations were not pursued any further with this approach.

The more recent black-hole perturbation approach
obtains high-order (in some cases exact) expressions
for the influence of radiation reaction on the orbital phase
which are valid in the limit of extreme mass ratio.
The basis of this approach is the perturbation of
known, exact solutions of the Einstein equations
(the Schwarzschild and Kerr spacetimes) with a test body
using the Teukolsky equation~\cite{Teukolsky73} or an equivalent. 
During the last several years, 
analytical techniques for post-Newtonian expansion
in the context of the black-hole perturbation approach
have been developed to very high orders in $v/c$
(for recent reviews, see~\cite{Supple}).
However, most black-hole perturbation papers treat the test body 
as a nonspinning point particle with a
$\delta$-function energy-momentum tensor,
and thus do not give results for the case of two spinning bodies.

Recently, the black-hole perturbation approach has been extended
to the case of two spinning bodies~\cite{MST96,TMSS96}.
In~\cite{MST96}, Mino, Shibata, and Tanaka calculated the
gravitational wave forms and radiation reaction
of a spinning particle falling into a Kerr black hole.
In~\cite{TMSS96}, Tanaka {\it et al.} obtained an expression for
the non-precessional 2.5PN spin-orbit contribution
to the secular phase evolution of a binary composed of
a spinning test particle in quasicircular orbit around a Kerr black hole.
These results were obtained using an energy-momentum tensor
for the test body which mimics the effects of an extended, spinning object 
but can be expressed in terms of a $\delta$-function for ease of calculation.
This ``spinning-particle'' $\delta$-function energy-momentum tensor
is based on the work of Dixon~\cite{Dixon79}.
We call it the MST tensor after Mino, Shibata, and Tanaka~\cite{MST96},
who distilled it into the compact form we will use.

In this paper, we use the MST energy-momentum tensor
for the first time in the curved-space, post-Newtonian approach
to derive new gravitational-wave generation results.
(Cho~\cite{Cho97}, in work parallel to our own,
has recently used a similar approach to re-derive the wave forms of
Kidder, Will, and Wiseman~\cite{KWW93} in a slightly different form.)
We reproduce (with a shorter calculation) the 1PN and 1.5PN 
spin-orbit corrections to the radiative multipoles derived in~\cite{Kidder95}.
We also derive all of the (previously unknown) 2PN non-precessional
spin-orbit corrections to the wave forms,
by calculating 2PN spin-orbit corrections to the radiative multipoles
and combining them with the well-known 1.5PN equations of motion
(in which there is no 2PN spin-orbit term).
Because of the harmonics of the orbital frequency involved,
there is no 2PN spin-orbit contribution to the
radiation reaction-induced secular phase evolution
of the wave forms (the most accurately measurable effect).

In the future, we plan to use the methods of this paper to calculate
all the non-precessional 2.5PN spin-orbit effects,
including the nonvanishing radiation reaction and resulting
secular evolution of the frequency and phase of the wave forms.
That secular evolution is likely to be quite important for data analysis.
Investigations by Tagoshi {\it et al.}~\cite{TSTS96},
comparing post-Newtonian expansions to exact numerical results
in the test-mass limit,
indicate that spin effects are important
for extraction of information from observed waves
at least up through 3PN order.

This paper is organized as follows.
In Sec.~\ref{s:T^ab} we present the MST energy-momentum tensor~\cite{MST96}
and review its properties.
In Sec.~\ref{s:pN} we review the post-Newtonian expansions
of basic variables used in our calculations.
Then in Sec.~\ref{s:multipoles} we calculate the STF radiative multipoles
needed to obtain the 2PN spin-orbit terms in the wave forms.
In Sec.~\ref{s:waveform} we evaluate
all the 2PN (non-precessional) spin-orbit terms
in the wave forms of a binary in quasicircular orbit
with spins parallel or antiparallel to the orbital angular momentum,
and in Sec.~\ref{s:summary} we briefly discuss their significance. 
In an Appendix we use our methods to derive the 1PN and 1.5PN
STF radiative multipoles, and compare with the results of
Refs.~\cite{KWW93,Kidder95}. 

Throughout this paper, we use units such that
Newton's gravitational constant and the speed of light equal unity.
We also use the tensor notation conventions of~\cite{BDI95,WW96}:
curved brackets $()$ on tensor indices to indicate symmetrization,
square brackets $[]$ to indicate antisymmetrization,
and angled brackets $\langle\rangle$ or the superscript STF
to indicate the symmetric trace-free part.
A capitalized superscript $L$ indicates a multi-index $i_1\cdots i_\ell$;
e.g., $I^L$ represents $I^{ijk}$ in the case $\ell=3$.
We also write outer products of vectors in shorthand;
e.g., $x^{ijk}=x^ix^jx^k$ and $x^L=x^{i_1}\cdots x^{i\ell}$.
Greek indices run from 0 to 3, and Latin indices from 1 to 3.

\section{Spinning particle energy-momentum tensor}
\label{s:T^ab}

Our starting point is the spinning particle energy-momentum tensor
given in terms of the Dirac $\delta$-function~\cite{MST96},
\begin{equation}
\label{e:T^ab}
T^{\alpha\beta}(x)=
\int d\tau\left\{p^{(\alpha}(x,\tau)\,u^{\beta)}(x,\tau)
\frac{\delta^{(4)}\bigl(x-z(\tau)\bigr)}{\sqrt{-g}}
-\nabla_\gamma\left[S^{\gamma(\alpha}(x,\tau)u^{\beta)}(x,\tau)
\frac{\delta^{(4)}\bigl(x-z(\tau)\bigr)}{\sqrt{-g}}\right]\right\}.
\end{equation} 
Here $z^\mu(\tau)$ is the worldline of the particle,
$u^\mu(\tau)=dz^\mu/d\tau$, $p^\mu(\tau)$ is the particle's linear momentum,
and $S^{\mu\nu}(\tau)$ is an antisymmetric tensor representing 
the particle's spin angular momentum.
We focus only on spin-orbit interactions,
i.e.\ discard all terms higher than first order in spin.
In this case $\tau$ becomes the particle's proper time
and $u^\mu$ becomes its four-velocity
[see Eq.~(2.4) of Ref.~\cite{MST96}].

The bitensors $p^\alpha(x,\tau)$, $u^\alpha(x,\tau)$, and
$S^{\alpha\beta}(x,\tau)$ are spacetime extensions of
$p^\mu$, $u^\mu$, and $S^{\mu\nu}$
away from the particle's worldline,\footnote{
We use indices $\alpha$, $\beta$, \ldots\ to denote quantities
associated with the field point $x$,
and $\mu$, $\nu$, \ldots\ to denote those associated
with the worldline $z(\tau)$.
}
defined by 
\begin{mathletters}
\begin{eqnarray}
p^\alpha(x,\tau)&=&\bar{g}^\alpha_{\ \mu}\bigl(x,z(\tau)\bigr)p^\mu(\tau),\\
u^\alpha(x,\tau)&=&\bar{g}^\alpha_{\ \mu}\bigl(x,z(\tau)\bigr)u^\mu(\tau),\\
S^{\alpha\beta}(x,\tau)&=&\bar{g}^\alpha_{\ \mu}\bigl(x,z(\tau)\bigr)
\bar{g}^\beta_{\ \nu}\bigl(x,z(\tau)\bigr)S^{\mu\nu}(\tau).
\end{eqnarray}
\end{mathletters}
Here $\bar{g}^\alpha_{\ \mu}(x,z)$ is a bitensor of parallel displacement 
with the properties
\begin{mathletters}
\begin{eqnarray}
\lim_{\scriptstyle x\to z}
\bar{g}^\alpha_{\ \mu}\bigl(x,z(\tau)\bigr)&=&\delta^\alpha_\mu,\\
\lim_{\scriptstyle x\to z}
\nabla_\beta\,\bar{g}^\alpha_{\ \mu}\bigl(x,z(\tau)\bigr)&=&0.
\end{eqnarray}
\end{mathletters}

The definition of $S^{\mu\nu}$ is arbitrary up to the choice of a
spin supplementary condition (the analog of a gauge condition).
We use
\begin{equation}
\label{e:SSC}
S^{\mu\nu}u_\mu = 0. 
\end{equation}
Note that in post-Newtonian theory 
at least three spin supplementary conditions are in common use.
We choose~(\ref{e:SSC}) because it makes our radiative multipoles
consistent with the standard post-Newtonian equations of motion, 
thus simplifying the calculations 
(cf.\ Ref.~\cite{Kidder95}, Appendix A). 
We introduce a spin vector $S^{\mu}$
which is related to the spin tensor by
\begin{mathletters}
\begin{eqnarray}
\label{e:S^mn}
S^{\mu\nu}&=&\epsilon^{\mu\nu\rho\sigma}u_\rho S_\sigma, \\
\label{e:SSC2}
S^\mu u_\mu&=&0, 
\end{eqnarray}
\end{mathletters}
where $\epsilon^{\mu\nu\rho\sigma}$ is
the Levi-Civita tensor.\footnote{
We use the convention defined in Eq.~(8.10) of Ref.~\cite{MTW}.} 
The spin supplementary condition is identically satisfied by~(\ref{e:S^mn}).
On the other hand, we need to impose the condition~(\ref{e:SSC2}) on $S^\mu$
to fix the one remaining degree of freedom $S^0$.

For later convenience, we separate
the MST energy-momentum tensor~(\ref{e:T^ab})
into the usual point-particle piece (the first term)
plus a spin-orbit piece, 
\begin{equation}
\label{e:T^ab_SO}
T^{\alpha\beta}_{\rm (SO)}(x)
=-\int d\tau\,\nabla_\gamma\left[S^{\gamma(\alpha}u^{\beta)}
\frac{\delta^{(4)}\bigl(x-z(\tau)\bigr)}{\sqrt{-g}}\right].
\end{equation}
When evaluating the radiative multipoles of a system of masses,
we encounter integrals of the form
\begin{equation}
\int d^3x\,F^L(x)T^{\alpha\beta}(x).
\end{equation}
The spin-orbit contribution to this expression can be evaluated
by substituting~(\ref{e:T^ab_SO}), using Leibniz' rule to rewrite the integral,
and discarding the spatial integral of a three-divergence.
The result for a many-body system is
\begin{equation}
\label{e:THI}
\int d^3x\,F^L(x)T^{\alpha\beta}_{\rm (SO)}(x)=\sum_A\biggl[
S_A^{\gamma(\alpha}v_A^{\beta)}\frac{\partial_\gamma F^L}{\sqrt{-g}}
-\partial_0\left(S_A^{0(\alpha}v_A^{\beta)}\frac{F^L}{\sqrt{-g}}\right)
-\left(\Gamma^\gamma_{\ \gamma\delta}S_A^{\delta(\alpha}v_A^{\beta)}
+S_A^{\gamma(\alpha}\Gamma^{\beta)}_{\ \gamma\delta}v_A^\delta\right)
\frac{F^L}{\sqrt{-g}}\biggr],
\end{equation}
where $A$ labels the bodies, $v^\alpha=u^\alpha/u^0$,
and $\partial_\gamma$ is shorthand for $\partial/\partial x^\gamma$
evaluated at $x=x_A$.

\section{Post-Newtonian expansions of basic variables}
\label{s:pN}

We now switch from fully covariant expressions
to post-Newtonian expansions in harmonic coordinates. 
Spatial indices on the right hand sides of the equations in this section 
can be raised and lowered freely with the Kronecker $\delta$.
We use the expansion parameter $\epsilon$ which is related to the orbital 
variables by $\epsilon \sim M/r \sim v^2$,
where $M$ is the total mass of the system, $r$ is the orbital separation,
and $v$ the orbital velocity.
We assume the bodies are compact, i.e.\ each body's spin has magnitude
$|\bbox{S}_A|\sim\chi m_A^2$, where $\chi$ is of order unity
(see~\cite{Kidder95} for further discussion).

When evaluating the post-Newtonian expansions of basic variables
in this section and the radiative multipoles in Sec.~\ref{s:multipoles},
we encounter divergent expressions---in our case, self-interaction terms.
Such divergences are inevitable when using any $\delta$-function source,
and we follow previous authors in discarding them
(see the discussion at the end of Ref.~\cite{BDI95}, Sec.~II).
We do not claim any rigorous justification for doing so;
however, since it is asserted in the non-spinning case~\cite{BDI95}
that this procedure can be justified to $O(\epsilon^2)$, 
and since we consider corrections only up to $O(\epsilon)$
beyond lowest order spin effects,
we expect that the formal use of the $\delta$-function 
is justified to the same degree as in the non-spinning case.
Informally, we note that
the usual post-Newtonian equations of motion for spinning bodies
can be obtained by taking the divergence of the
MST energy-momentum tensor~(\ref{e:T^ab}) and discarding self-interaction
divergences~\cite{MST96}.

The metric components in harmonic coordinates are well known~\cite{WW96} as
\begin{mathletters}
\label{e:g_ij}
\begin{eqnarray}
g_{00}&=&-[1-2U+O(\epsilon^2)],\\
g_{i0}&=&O(\epsilon^{3/2}),\\
g_{ij}&=&\delta_{ij}[1+2U+O(\epsilon^2)],\\
\sqrt{-g}&=&1+2U+O(\epsilon^{3/2}),
\end{eqnarray}
\end{mathletters}
where only the lowest-order expression for the potential $U$ is needed,
\begin{equation}
U(\bbox{x})=\sum_A\frac{m_A}{|\bbox{x}-\bbox{x}_A|} + O(\epsilon^2).
\end{equation}
By differentiating~(\ref{e:g_ij}),
we find the dominant Christoffel symbols
\begin{mathletters}
\begin{eqnarray}
\Gamma^0_{i0} &=& \Gamma^i_{00} = -a^i ,\\
\Gamma^i_{jk} &=& \delta^{ij}a^k + \delta^{ik}a^j - \delta^{jk} a^i ,
\end{eqnarray}
\end{mathletters}
where $a^i=\partial_iU$.
All others are of higher post-Newtonian order,
and can be neglected for the purposes of this paper. 
The metric components~(\ref{e:g_ij}), together with the condition
$u^\mu u_\mu = -1$, give us the expansion of the four-velocity
\begin{mathletters}
\label{e:u^m}
\begin{eqnarray}
u^0=&&1+\left({v^2\over 2}+U\right)
+O(\epsilon^2),\\
u^i=&&v^i\left[1+\left({v^2\over 2}+U\right)+
O(\epsilon^{2}) \right].
\end{eqnarray}
\end{mathletters}

We express the components of the spin tensor
in terms of the spatial components of the spin vector by combining 
(\ref{e:S^mn}), (\ref{e:SSC2}), (\ref{e:g_ij}), and~(\ref{e:u^m}) to obtain
\begin{mathletters}
\label{e:S^ij}
\begin{eqnarray}
S^{i0}&=&(\bbox{v}\times\bbox{S})^i+O(\epsilon^{3/2}),\\
S^{ij}&=&\epsilon^{ijk}
\left[(1+\frac{1}{2}v^2-U)S^k-(\bbox{v}\cdot\bbox{S})v^k\right]\nonumber\\
&&+O(\epsilon^2),
\end{eqnarray}
\end{mathletters}
where $\epsilon^{ijk}$ is from here on used to indicate
the antisymmetric symbol $[ijk]$.
Substituting~(\ref{e:S^ij}) back into~(\ref{e:T^ab_SO}),
we find that the post-Newtonian orders of the components of
$T^{\alpha\beta}_{\rm (SO)}$ are
\begin{mathletters}
\label{e:Torder}
\begin{eqnarray}
T^{00}_{\rm (SO)}\sim T^{ij}_{\rm (SO)}&\sim&
m\frac{mv}{r}\sim m\times O(\epsilon^{3/2}),\\
T^{i0}_{\rm (SO)}&\sim&m\frac{m}{r}\sim m\times O(\epsilon),
\end{eqnarray}
\end{mathletters}
where $m$ is the mass of either body.
This contrasts with the point-mass order counting,
\begin{mathletters}
\begin{eqnarray}
T^{00}_{\rm (PM)}&\sim&m,\\
T^{i0}_{\rm (PM)}&\sim&m\times O(\epsilon^{1/2}),\\
T^{ij}_{\rm (PM)}&\sim&m\times O(\epsilon).
\end{eqnarray}
\end{mathletters}
We also note that the second and third terms
on the right hand side of~(\ref{e:THI})
are $O(\epsilon)$ with respect to the first if $F^L$ is an outer product
of position vectors (as is the case when computing multipoles).

\section{STF radiative multipoles}
\label{s:multipoles}

In this section we calculate
the symmetric, trace-free (STF) radiative multipoles
necessary to obtain the 2PN spin-orbit contributions to the wave form.
In Sec.~\ref{ss:circular} we specialize to the case of non-precessing orbits.

The STF radiative multipoles are given to $O(\epsilon^{5/2})$
by Blanchet [Eq.~(4.3) of Ref.~\cite{Blanchet96}].
However, we only need the $O(\epsilon)$ expressions
\begin{eqnarray}
\label{e:I^L}
I^L(t)&=&
\int d^3x\,\hat{x}^L\left(T^{00}+T^{ii}\right)
-\frac{4(2\ell+1)}{(\ell+1)(2\ell+3)}\frac{d}{dt}
\int d^3x\,\hat{x}^{La}T^{0a} \nonumber\\
&&
+{1 \over 2(2\ell+3)}{d^2\over dt^2}
\int d^3 x\,{\hat{x}_L |\bbox{x}|^2}\left(T^{00}+T^{ii}\right)+O(\epsilon^2),\\
\label{e:J^L}
J^L(t)&=&
\epsilon^{ab\langle i}\biggl[
\int d^3x\,\hat{x}^{L-1\rangle a}(1+4U)T^{0b}
-\frac{2\ell+1}{(\ell+2)(2\ell+3)}\frac{d}{dt}
\int d^3x\,\hat{x}^{L-1\rangle ac}T^{bc}\nonumber\\
&&+\frac{1}{2(2\ell+3)}\frac{d^2}{dt^2}
\int d^3x\,\hat{x}^{L-1\rangle a}|\bbox{x}|^2T^{0b}\biggr]
+O(\epsilon^2), 
\end{eqnarray}
where $\hat{x}^L$ denotes the symmetric trace-free part of $x^L$. 
We also define $I^L_{\rm (SO)}$ and $J^L_{\rm (SO)}$ by substituting 
$T^{\mu\nu}_{\rm (SO)}$ for $T^{\mu\nu}$ in (\ref{e:I^L}) and (\ref{e:J^L}). 
In~(\ref{e:I^L}) and (\ref{e:J^L}) we have discarded self-interaction terms,
which are always divergent when using a $\delta$-function source.
We have also discarded terms involving only gravitational potentials
(referred to as ``non-compact'' terms in~\cite{BDI95}),
whose spin-orbit contributions do not appear until higher post-Newtonian orders
than considered in this paper.

Spin-orbit corrections to the multipoles and wave form
follow an order-counting scheme different from the usual point-mass terms.
Substituting~(\ref{e:Torder}) into~(\ref{e:I^L}) and (\ref{e:J^L}),
we find that the lowest order spin-orbit correction to a multipole
appears in the current quadrupole $J^{ij}_{\rm (SO)}$~\cite{Kidder95}. 
This term contributes to the wave-form amplitude at 1PN order,
but because it is a subharmonic of the dominant (mass quadrupole) radiation, 
it does not contribute to the radiation reaction until 1.5PN order.
The next-order effects appear in the current octupole $J^{ijk}_{\rm (SO)}$
and the mass quadrupole $I^{ij}_{\rm (SO)}$,
and contribute to the wave forms and radiation reaction at 1.5PN order.
(These are the terms given in~\cite{KWW93}; we evaluate them with our methods
in the appendix.)
Following this progression, the 2PN wave forms require evaluation of 
$J^{ijk\ell}_{\rm (SO)}$ and $I^{ijk}_{\rm (SO)}$ to lowest order,
and of $J^{ij}_{\rm (SO)}$ to $O(\epsilon)$ beyond lowest order. 

\subsection{$N$-body case}
\label{ss:sumform}

We first evaluate the spin-orbit contributions to the 
multipole integrals~(\ref{e:I^L}) and (\ref{e:J^L}) 
as sums over $N$ bodies.

The expression for the current hexadecapole is the easiest to evaluate.
It is needed only to lowest order and thus involves only the first term
in~(\ref{e:J^L}),
\begin{equation}
J^{ijk\ell}_{\rm (SO)}
=\epsilon^{ab\langle i}\int d^3x\,\hat{x}^{jk\ell\rangle a}T^{0b}_{\rm (SO)},
\end{equation}
which is straightforwardly obtained from the first term of~(\ref{e:THI}) as
\begin{equation}
\label{e:J^ijkl,sum}
J^{ijk\ell}_{\rm (SO)}=
\frac{5}{2}\sum_A \left(S_A^i x_A^{jk\ell}\right)^{\rm STF}.
\end{equation}

The mass octupole is also needed only to lowest order,
\begin{equation}
I^{ijk}_{\rm (SO)}=\int d^3x\,x^{ijk}
\left(T^{00}_{\rm (SO)}+T^{aa}_{\rm (SO)}\right)
-\frac{7}{9}\frac{d}{dt}\int d^3x\,x^{ijka}T^{0a}_{\rm (SO)}.
\end{equation}
Again, using the first term of~(\ref{e:THI})
it is straightforward to evaluate the integrals.
When evaluating the time derivative in the second term
we neglect time derivatives of the spins (i.e., precession).
We could do this even if considering spin precession, because
those derivatives appear $O(\epsilon)$ beyond the spins themselves
[cf.\ Eqs.~(F18,F19) of Ref.~\cite{WW96}, where due to a typographical error
a factor of $(M/r)^3$ was omitted from in front of the last term in each
equation].\footnote{
Although we do not consider spin precession in this paper,
the $N$-body and two-body multipoles we present are
(instantaneously) valid even in the precessing case.
One simply has to put in the spin vectors as (slowly varying)
functions of time---which is easier said than done.
}
We are left with
\begin{equation}
\label{e:I^ijk,sum}
I^{ijk}_{\rm (SO)}=\sum_A
\left[\frac{9}{2}(\bbox{v}_A\times\bbox{S}_A)^i x_A^{jk}
-3(\bbox{x}_A\times\bbox{S}_A)^i x_A^j v_A^k\right]^{\rm STF}
\end{equation}

Because the two-index current moment is needed to $O(\epsilon)$
we must keep all three terms in~(\ref{e:J^L}),
\begin{equation}
J^{ij}_{\rm (SO)}=\epsilon^{ab\langle i}\biggl[
\int d^3x\,x^{j\rangle a}(1+4U)T^{0b}_{\rm (SO)}
-\frac{5}{28}\frac{d}{dt}
\int d^3x\,x^{j\rangle ac}T^{bc}_{\rm (SO)}
+\frac{1}{14}\frac{d^2}{dt^2}
\int d^3x\,x^{j\rangle a}T^{0b}_{\rm (SO)}\biggr].
\end{equation}
Again, time derivatives of the spins appear only in higher-order terms
and may be discarded at this order.
Carefully evaluating the integrals according to~(\ref{e:THI}),
we obtain
\begin{eqnarray}
\label{e:J^ij,sum}
J^{ij}_{\rm (SO)}&=&\sum_A\biggl[
{3\over 2}x_A^i S_A^j+
{1\over 14} \bbox{v}_A\cdot\bbox{x}_A v_A^{i} S_A^{j} + 
{2\over 7} \bbox{S}_A\cdot\bbox{x}_A v_A^{ij} + 
{11\over 28} \bbox{x}_A\cdot\bbox{x}_A  a_A^{i} S_A^{j}- 
{17\over 7}  \bbox{S}_A\cdot\bbox{v}_A x_A^{i} v_A^{j} \nonumber\\
&&
+{1\over 7} \bbox{a}_A\cdot\bbox{S}_A x_A^{ij}
-{17\over 14} \bbox{S}_A\cdot\bbox{x}_A x_A^{i} a_A^{j} 
+\big({3\over 2} U_A + 
{43\over 28} \bbox{v}_A\cdot\bbox{v}_A + 
{11\over 14} \bbox{a}_A\cdot\bbox{x}_A \bigr)x_A^{i} S_A^{j}
\biggr]^{\rm STF}
\end{eqnarray}

In addition to the spin-orbit multipoles,
we need the lowest order contributions to $I^{ijk}$ and $J^{ij}$
from the usual point-mass energy-momentum tensor. 
These multipoles are given by
\begin{eqnarray}
\label{e:PMmulti}
I^{ijk}_{\rm (PM)}&=&\sum_Am_Ax_A^{\langle ijk\rangle},\\
J^{ij}_{\rm (PM)}&=&
\sum_Am_A(\bbox{x}_A\times\bbox{v}_A)^{\langle i}x_A^{j\rangle}
\end{eqnarray}
and contribute to the wave forms at 0.5PN order. 

\subsection{Two-body case}
\label{ss:twobody}

We now specialize to the case of two bodies of mass $m_1$ and $m_2$
in an arbitrary (possibly precessing) orbit,
and express the multipoles in terms of the relative coordinate $\bbox{x}$
(whose origin is at body 2).

It is convenient to use the mass parameters
\begin{mathletters}
\begin{eqnarray}
M&=&m_1+m_2,\\
\eta&=&m_1m_2/M^2,\\
\Delta&=&(m_1-m_2)/M.
\end{eqnarray}
\end{mathletters}
It is also convenient to use the dimensionless, symmetrized spin parameters
introduced by Will and Wiseman~\cite{WW96},
\begin{mathletters}
\begin{eqnarray}
\bbox{\chi}_{a}&=&\frac{1}{2}
\left(\frac{\bbox{S}_1}{m_1^2}-\frac{\bbox{S}_2}{m_2^2}\right), \\
\bbox{\chi}_{s}&=&\frac{1}{2}
\left(\frac{\bbox{S}_1}{m_1^2}+\frac{\bbox{S}_2}{m_2^2}\right),
\end{eqnarray}
\end{mathletters}
We eliminate the potentials and accelerations in~(\ref{e:J^ij,sum}) with
the well-known Newtonian expressions
\begin{eqnarray}
U_A&=&U(\bbox{x}_A)=\frac{m_B}{r},\\
\bbox{a}_A&=&\frac{-M}{r^3}\bbox{x}_A,
\end{eqnarray}
where $B\neq A$ and $r=|\bbox{x}|$.

The spins of the bodies introduce a correction
to the relation between $\bbox{x}_1$, $\bbox{x}_2$,
and the relative coordinate $\bbox{x}$, 
\begin{mathletters}
\begin{eqnarray}
\label{e:com}
\bbox{x}_1&=&\bbox{x}\left[{m_2\over M}+{1\over 2}\eta
\Delta\Bigl(v^2-{M\over r}\Bigr)\right]
-M\eta\bbox{v}\times(\bbox{\chi}_a+\Delta\bbox{\chi}_s), 
\\
\bbox{x}_2&=&\bbox{x}\left[-{m_1\over M}+{1\over 2}\eta
\Delta\Bigl(v^2-{M\over r}\Bigr)\right]
-M\eta\bbox{v}\times(\bbox{\chi}_a+\Delta\bbox{\chi}_s).
\end{eqnarray}
\end{mathletters}
(Compare Eq.~(3.13) of Ref.~\cite{Kidder95} and
Eq.~(F11) of Ref.~\cite{WW96}, where the missing factor of $\eta$
in the latter is a typographical error.)
This correction is 1.5PN order; therefore it enters our 2PN calculation
through contributions from the 0.5PN (point-mass)
multipoles~(\ref{e:PMmulti}). 

Applying the transformation~(\ref{e:com}) and including the contributions
from the point-mass multipoles, we find the two-body forms
of~(\ref{e:J^ijkl,sum}), (\ref{e:I^ijk,sum}), and~(\ref{e:J^ij,sum}) to be
\begin{equation}
\label{e:J^ijkl,two}
J^{ijk\ell}=\frac{5}{2}M^2\eta^2
(\chi_a-\Delta\chi_s)^{\langle i}x^{jk\ell\rangle},
\end{equation}
\begin{equation}
\label{e:I^ijk,two}
I^{ijk}=
-M\eta\Delta x^{\langle ijk \rangle}
+M^2\eta^2\biggl\{
\frac{3}{2}[\bbox{v}\times(\bbox{\chi}_a-5\Delta\bbox{\chi}_s)]^ix^{jk}
-3[\bbox{x}\times(\bbox{\chi}_a-\Delta\bbox{\chi}_s)]^ix^{jk}
\biggr\}^{\rm STF},
\end{equation}
\begin{eqnarray}
\label{e:J^ij,two}
J^{ij}&=&
-\mu\Delta (\bbox{x}\times\bbox{v})^{\langle i}x^{j\rangle}
+{3\over 2}M^2\eta(\chi_a+\Delta \chi_s)^{\langle i}x^{j\rangle}
+{3\over 4}M^2\eta\Delta\left(v^2-{M\over r}\right)
(\Delta\chi_a+\chi_s)^{\langle i}x^{j\rangle}\nonumber\\
&&
+{1\over 28}M^2\eta^2\biggl\{\left[\left(23{M\over r}-13v^2\right)\chi_a
+\Delta\left(47{M\over r}-141v^2\right)\chi_s\right]^ix^j\nonumber\\
&&+2(\bbox{x}\cdot\bbox{v})(15\chi_a+13\Delta\chi_s)^iv^j
+2{M\over r^3}[\bbox{x}\cdot(29\bbox{\chi}_a-\Delta\bbox{\chi}_s)]x^{ij}
-4[\bbox{x}\cdot(5\bbox{\chi}_a+9\Delta\bbox{\chi}_s)]v^{ij}\nonumber\\
&&-4[\bbox{v}\cdot(3\bbox{\chi}_a-31\Delta\bbox{\chi}_s)]x^iv^j
\biggr\}^{\rm STF}.
\end{eqnarray}
The first terms in (\ref{e:I^ijk,two}) and (\ref{e:J^ij,two}) are 
the lowest order non-spin terms. 
The second term in~(\ref{e:J^ij,two}) is easily verified as identical
to the lowest order (1PN) spin-orbit contribution obtained by Kidder
[Eq.~(3.20a) of Ref.~\cite{Kidder95}].

\subsection{Quasicircular orbits}
\label{ss:circular}

We now specialize to the case where the two bodies orbit each other
in a circular trajectory, which adiabatically shrinks (inspirals)
under radiation reaction.
For spinning bodies, this is only possible when both spin vectors are
parallel or antiparallel to the orbital angular momentum,
eliminating spin-orbit precession.

We express our results in terms of the orthonormal vectors
$\bbox{n}=\bbox{x}/r$, $\bbox{\lambda}=\bbox{v}/v$,
and $\bbox{z}=\bbox{n}\times\bbox{\lambda}$
(parallel to all angular momenta).
The majority of the terms in~(\ref{e:J^ijkl,two}-\ref{e:J^ij,two}) vanish,
and we are left with the greatly simplified expressions
\begin{equation}
\label{e:J^ijkl,circ}
J^{ijk\ell}
=\frac{5}{2}M^2\eta^2r^3(\chi_a-\Delta\chi_s)n^{\langle ijk}z^{\ell\rangle},
\end{equation}
\begin{equation}
\label{e:I^ijk,circ}
I^{ijk}=
-M\eta\Delta r^3 n^{\langle ijk\rangle}
+\frac{3}{2}M^2\eta^2vr^2\left[(\chi_a-5\Delta\chi_s)n^{ijk}
+2(\chi_a-\Delta\chi_s)n^i\lambda^{jk}\right]^{\rm STF},
\end{equation}
\begin{equation}
\label{e:J^ij,circ}
J^{ij}=
-M\eta\Delta r^2 v n^{\langle i}z^{j \rangle}
+\frac{3}{2}M^2\eta r(\chi_a+\Delta\chi_s)n^{\langle i}z^{j\rangle}
+\frac{1}{14}M^3\eta^2(5\chi_a-47\Delta\chi_s)n^{\langle i}z^{j\rangle}.
\end{equation}

\section{Wave form}
\label{s:waveform}

Having evaluated the necessary radiative multipoles, we obtain the
gravitational wave form
\begin{equation}
\label{e:h^ij,gen}
h^{ij}=\frac{1}{R}\sum_{\ell=2}^\infty\biggl[
\frac{4}{\ell!}\stackrel{(\ell)\ \ \ \ \ }{I^{ijL-2}}N^{L-2}
+\frac{8\ell}{(\ell+1)!}\epsilon^{pq(i}
\stackrel{(\ell)\ \ \ \ \ \ }{J^{j)pL-2}}N^{qL-2}
\biggr]^{\rm TT},
\end{equation}
where TT denotes the transverse traceless projection,
$(\ell)$ denotes the $\ell$th time derivative,
and $\bbox{N}$ is the unit vector pointing toward the observer
[cf.\ Eq.~(E5a) of Ref.~\cite{WW96}].

We evaluate the time derivatives using the equation of motion
for circular orbits, $\bbox{a}=-\omega^2\bbox{x}$,
where the standard form of the post-Newtonian expansion is given by
\begin{equation}
\label{e:EOM}
\omega^2=\frac{M}{r^3}\biggl\{1-(3-\eta)\frac{M}{r}
-2[\Delta\chi_a+(1+\eta)\chi_s]\frac{Mv}{r}+O(\epsilon^2)\biggr\}
\end{equation}
[cf.\ Eqs.~(7.1) and~(F20) of Ref.~\cite{WW96}].
The 1.5PN spin-orbit term in the equation of motion
[the third term in~(\ref{e:EOM})]
contributes to the 2PN spin-orbit term in $h^{ij}$
through time derivatives of the 0.5PN point-mass multipoles
[cf.\ the first terms in (\ref{e:I^ijk,circ}) and (\ref{e:J^ij,circ})].
If we were to consider spin-orbit precession,
time derivatives of spin expressions appearing in the
multipoles~(\ref{e:J^ijkl,two})-(\ref{e:J^ij,two})
would also factor into~(\ref{e:h^ij,gen}).

Evaluating the time derivatives in~(\ref{e:h^ij,gen}),
using~(\ref{e:EOM}) and the identity $v=\omega r$ to write
$v$ and $M/r$ in terms of $M\omega$,
and collecting all terms of order $(M\omega)^2$,
we obtain
\begin{eqnarray}
\label{e:h^ij}
h^{ij}&=&{2M\eta\over R}(M\omega)^2\biggl[
(\Delta^2\chi_{a}+\Delta\chi_s)
\Bigl(\Bigl(-{14\over 3}n^{ij}+4\lambda^{ij}\Bigr)(\bbox{\lambda}\cdot\bbox{N})
-{28\over 3}n^{(i}\lambda^{j)}(\bbox{n}\cdot\bbox{N})\Bigr) \nonumber\\
&&
+\Bigl(2\chi_a-{8\over 3}\Delta^2\chi_a-{2\over 3}\Delta\chi_s\Bigr)
(\bbox{n}\times\bbox{N})^{(i}z^{j)} \biggr]^{\rm TT}\nonumber\\
&&
+{2M\eta^2\over R}(M\omega)^2\biggl[
\Bigl(19\chi_a+37\Delta\chi_s\Bigr)
\Bigl({1\over 6}(\bbox{N}\cdot\bbox{\lambda})n^{ij}
+{1\over 3}(\bbox{N}\cdot\bbox{n})n^{(i}\lambda^{j)} \Bigr) \nonumber\\
&&-4(\chi_{a}+\Delta\chi_s)(\bbox{N}\cdot\bbox{\lambda})\lambda^{ij}
+(\chi_{a}-\Delta\chi_s)\biggl\{
(\bbox{n}\times\bbox{N})^{(i}z^{j)}
\Bigl(-{7\over 6}-{10\over 3} (\bbox{N}\cdot\bbox{\lambda})^2+
{7\over 2}(\bbox{N}\cdot\bbox{n})^2\Bigr)\nonumber\\
&&-{20\over 3}(\bbox{N}\cdot\bbox{\lambda})(\bbox{N}\cdot\bbox{z})
(\bbox{n}\times\bbox{N})^{(i}\lambda^{j)}
+{7\over 3}(\bbox{N}\cdot\bbox{n})(\bbox{N}\cdot\bbox{z})
(\bbox{n}\times\bbox{N})^{(i}n^{j)} \nonumber\\
&&-{10\over 3}(\bbox{N}\cdot\bbox{n})(\bbox{N}\cdot\bbox{z})
(\bbox{\lambda}\times\bbox{N})^{(i}\lambda^{j)}
-{20\over 3}(\bbox{N}\cdot\bbox{n})(\bbox{N}\cdot\bbox{\lambda})
(\bbox{z}\times\bbox{N})^{(i}\lambda^{j)} \biggr\}\biggr]^{\rm TT}. 
\end{eqnarray}

Finally, we project out the amplitudes of the
$h_+$ and $h_\times$ polarizations.
This is done by taking
\begin{mathletters}
\begin{eqnarray}
h_+&=&\frac{1}{2}(p_ip_j-q_iq_j)h^{ij},\\
h_\times&=&\frac{1}{2}(p_iq_j+q_ip_j)h^{ij}.
\end{eqnarray}
\end{mathletters}
Note that the TT projection in~(\ref{e:h^ij,gen}) is subsumed in this operation
(cf.\ Sec.~VIIA of~\cite{WW96}).
The standard convention~\cite{WW96} is to use the unit triad composed of
$\bbox{N}$ (the direction to the observer),
$\bbox{p}$ (pointing from the descending node of the orbit
to the ascending node), 
and $\bbox{q}=\bbox{N}\times\bbox{p}$.
The orbital phase $\psi$ is measured from the ascending node,
and the orbital inclination angle $i$ is given by
$\cos{i}=\bbox{z}\cdot\bbox{N}$. 
Thus we have
\begin{mathletters}
\begin{eqnarray}
\bbox{n}&=&\bbox{p} \cos\psi
+(\bbox{q} \cos i+\bbox{N} \sin i)\sin\psi,\\ 
\bbox{\lambda}&=&-\bbox{p}\sin\psi
+(\bbox{q} \cos i+\bbox{N} \sin i)\cos\psi,\\
\bbox{z}&=&-\bbox{q}\sin i+\bbox{N}\cos i.
\end{eqnarray}
\end{mathletters} 

Following~\cite{BIWW96}, we organize the amplitude contributions
of the wave-form polarizations according to post-Newtonian order
and physical origin as
\begin{equation}
h_{+,\times}=\frac{2M\eta}{R}x\biggl[
H_{+,\times}^{(0)}+\cdots+x^{3/2}H_{+,\times}^{(3/2)}
+x^{3/2}H_{+,\times}^{\rm(3/2,SO)}
+x^2H_{+,\times}^{(2)}+x^2H_{+,\times}^{\rm(2,SO)}
+x^2H_{+,\times}^{\rm(2,SS)}+\cdots\biggr],
\end{equation}
where $x=(M\omega)^{2/3}$.
The 2PN spin-orbit contributions to the wave-form polarizations are given by
\begin{mathletters}
\label{e:results}
\begin{eqnarray}
H_+^{\rm (2,SO)}&=&
\Bigl\{\frac{-1}{24}\Bigl[\left(109 + 15c^2 \right)\chi _{a}
+ 7\left(1 + 3 c^2 \right)\Delta \chi _{s}\Bigr]\eta
+ {1\over 4}(1 + c^2 )(\chi _a + \Delta\chi_s)
\Bigr\}\sin i \,\cos \psi \nonumber\\
&&
+ \Bigl\{\frac{9}{8}\Bigl[\left(11 + 5 c^2 \right)\chi _{a}
+ \left(1 + 7 c^2 \right)\Delta \chi _{s}\Bigr]\eta
-{9\over 4} (1 + c^2 )(\chi_a+\Delta\chi_s) \Bigr\} 
\sin i \,\cos 3\psi 
\end{eqnarray}
\begin{eqnarray}
H_\times^{\rm (2,SO)}&=&
\Bigl\{\frac{-1}{24}\Bigl[\left(127 - 3 c^2 \right)\chi _{a}
+ \left(25 + 3 c^2 \right)\Delta \chi _{s}\Bigr]\eta
+ \frac {1}{2} (\chi_a + \Delta \chi_s)
\Bigr\} c \,\sin i \,\sin \psi \nonumber\\
&&
+ \Bigl\{\frac{9}{8}\Bigl[\left(19 - 3 c^2 \right)\chi _{a}
+ \left(5 + 3 c^2 \right)\Delta \chi _{s}\Bigr]\eta
 - \frac {9}{2} (\chi_a + \Delta \chi_s)
\Bigr\}c \,\sin i \,\sin 3\psi ,
\end{eqnarray}
\end{mathletters}
where $c\equiv \cos i$.

It is clear that these 2PN contributions to the wave-form amplitudes
do not contribute to the radiation reaction at 2PN order
because the harmonics ($\omega$ and $3\omega$) average to zero
when beat against the ``Newtonian'' terms $H_{+,\times}^{(0)}$
at frequency $2\omega$ [cf.\ Eqs.~(3,4) of Ref.~\cite{BIWW96}].

\section{Summary}
\label{s:summary}

We have calculated all the non-precessional 2PN spin-orbit effects
on the gravitational wave forms of compact bodies in quasicircular orbit,
and have shown that there is no spin-orbit radiation reaction
effect at 2PN order.
Our calculation was greatly simplified
over previous spinning-body post-Newtonian
efforts~\cite{KWW93,Kidder95} by the use of a $\delta$-function
energy-momentum tensor for spinning particles.
We have presented the wave-form polarizations in ``ready-to-use'' form
(cf.~\cite{BIWW96}).

Note that terms of $O(\eta)$ contribute significantly to~(\ref{e:results}). 
These are the terms that could not have been obtained by the 
black-hole perturbation approach. 
Their presence leads us to expect that the 2.5PN radiation reaction
will also contain significant terms of $O(\eta)$ 
which are not found in~\cite{TMSS96,SSTT95,TSTS96}. 

In this paper, we treated the bodies only to linear order in their spins
(i.e., considered only spin-orbit effects).
Spin-spin effects can be treated by a more complicated calculation
using the MST energy-momentum tensor.
However, because spin-spin effects appear at 2PN order,
consistency would require that one also include the effects
of the bodies' quadrupole moments~\cite{Poisson97}.

\acknowledgments

We thank Kip Thorne for helpful comments on the manuscript.
BJO thanks Alan Wiseman for many stimulating discussions.
BJO was supported in part by NSF Grants AST-9417371 and INT-9417348,
NASA Grants NAGW-4268 and NAG5-4351, and by an NSF Graduate Fellowship. 
HT was supported by the Japanese Society for the Promotion of Science. 
AO was supported in part by Monbusho Grant-in-Aid for 
Scientific Research No.\ 09440106. 

\appendix
\section*{}
\label{s:lower}

In this appendix, we derive the 1.5PN spin-orbit multipoles 
$I^{ij}_{\rm (SO)}$ and $J^{ijk}_{\rm (SO)}$ obtained in~\cite{Kidder95}.

The general expressions are quite simple,
requiring~(\ref{e:I^L}) and~(\ref{e:J^L}) only to lowest order,
\begin{eqnarray}
I^{ij}_{\rm (SO)}&=&
\int d^3x\,x^{ij}\left(T^{00}_{\rm (SO)}+T^{aa}_{\rm (SO)}\right)\nonumber\\
&&-\frac{4}{3}\frac{d}{dt}\int d^3x\,x^{ijka}T^{0a}_{\rm (SO)},\\
J^{ijk}_{\rm (SO)}&=&
\epsilon^{ab\langle i}\int d^3x\,x^{jk\rangle a}T^{0b}_{\rm (SO)}.
\end{eqnarray}
Using~(\ref{e:THI}) and~(\ref{e:S^ij}) only to lowest order,
it is straightforward to evaluate these expressions in the $N$-body case as
\begin{eqnarray}
I^{ij}_{\rm (SO)}&=&\sum \bigl[4x_A^i(\bbox{v}_A\times\bbox{S}_A)^j\nonumber\\
&&-{4\over 3}{d\over dt}
\left\{x_A^i(\bbox{x}_A\times\bbox{S}_A)^j\right\}\bigr]^{\rm STF}, \\
J^{ijk}_{\rm (SO)}&=&\sum 2[x_A^{ij}S_A^k]^{\rm STF}. 
\end{eqnarray}

Using the transformation~(\ref{e:com}) to the relative coordinate, 
we find the two-body forms of the multipoles to be
\begin{eqnarray}
I^{ij}_{\rm (SO)}&=&{8\over 3}M^2\eta^2
[2x^i(\bbox{v}\times\bbox{\chi}_s)^j
-v^i(\bbox{x}\times\bbox{\chi}_s)^j]^{\rm STF}, \\
J^{ijk}_{\rm (SO)}&=&4 M^2\eta^2[x^{ij}\chi_s^k]^{\rm STF}.
\end{eqnarray}
These results agree with those obtained using the fluid body
energy-momentum tensor~\cite{Kidder95,WW96}.

\end{document}